\documentclass[aps,prd,nofootinbib,preprintnumbers,eqsecnum]{revtex4}
\usepackage{amsmath}
\usepackage{amssymb, amsmath}
\usepackage{amsfonts}
\usepackage{graphicx}
\usepackage{bm}

\usepackage{color}

\def\mc{\mathcal}
\def\mr{\mathrm}
\def\pd{\partial}

\newcommand{\diff}[3]{\frac{\mr{d}^{#1} #2}{\mr{d}#3^{#1}}}

\begin{document}

\preprint{YITP-12-74}

\title{Hawking-Moss instanton in nonlinear massive gravity}

\date{\today}
\author{Ying-li Zhang\footnote{E-mail
 address: yingli{}@{}yukawa.kyoto-u.ac.jp},
Ryo Saito\footnote{E-mail
 address: rsaito{}@{}yukawa.kyoto-u.ac.jp} and Misao Sasaki\footnote{E-mail
 address: misao{}@{}yukawa.kyoto-u.ac.jp}
 }

\affiliation{Yukawa Institute for Theoretical Physics, Kyoto
University, Kyoto 606-8502, Japan }

\begin{abstract}
As a first step toward understanding a lanscape of vacua in a theory of
 non-linear massive gravity,
we consider a landscape of a single scalar field and study tunneling
between a pair of adjacent vacua. We study the Hawking-Moss (HM)
instanton that sits at a local maximum of the potential, and
evaluate the dependence of the tunneling rate on the parameters of
the theory. It is found that provided with the same physical HM
Hubble parameter $H_{HM}$, depending on the values of parameters
$\alpha_3$ and $\alpha_4$ in the action (\ref{eq:actiong}), the
corresponding tunneling rate can be either enhanced or suppressed
when compared to the one in the context of General Relativity (GR).
Furthermore, we find the constraint on the ratio of the physical
Hubble parameter to the fiducial one, which constrains the form of
potential. This result is in sharp contrast to GR where there is no
bound on the minimum value of the potential.
\end{abstract}

\maketitle

\section{Introduction}

It is of great interest to promote Einstein's gravity theory
 to a massive one \cite{Fierz:1939}.
However, it has been shown that Lorentz-invariant massive gravity theories
would suffer from the Boulware-Deser (BD) ghost
instability~\cite{Boulware:1972, Creminelli:2005qk}
(For a review, see e.g. Ref.~\cite{Rubakov:2008, Hinterbichler:2012}).
Recently, a non-linear construction of a Lorentz-invariant massive gravity
theory has been proposed in Refs. \cite{Rham:2010, Rham:2011PRL, Hassan:2011vm},
where the nonlinear terms are designed so that the BD ghost is removed by
using the constraint equations \cite{Hassan:2012, Hassan:2011tf}.
Many researches have been made to explore its physical consequences since this
breakthrough.

One of the most interesting results among these explorations is that
such a massive gravity theory allows self-accelerating solutions
(e.g. for Minkowski fiducial metric, see~\cite{D'Amico:2011jj,
Gumrukcuoglu:2011open, Kobayashi:2012, Gratia:2012, Koyama:2012prda,
Koyama:2012prl}; for deSitter fiducial metric,
see~\cite{Langlois:2012hk}; for a more general case where fiducial
metric respects only isotropy, see~\cite{ Motohashi:2012jd}). While
cosmological solutions in usual massive gravity theories are known
to
 suffer from appearance of a ghost within $5$ degrees of freedom
of the massive graviton \cite{Higuchi:1986py, Grisa:2009yy, Fasiello:2012rw},
it has been proved that perturbations on the solution found
 in Ref. \cite{Gumrukcuoglu:2011open} do not have massive degrees of freedom
and evade the ghost instability at linear order
\cite{Gumrukcuoglu:2011perturb}. \footnote{See also the recent
discussion on the ghost instability at nonlinear level
\cite{Koyama:2011jhep, DeFelice:2012mx, Koyama:2012vector}.}

Though the mass of graviton may successfully explain the current acceleration,
 it is still to be explained why the cosmological constant is so small
while large quantum corrections are expected.
\footnote{Note that the graviton mass does not receive a large quantum
 correction because the other parts, where the general coordinate
invariance symmetry is respected, do not induce a correction to the mass terms.}
In spite of many attempts,
this problem, so called ``the cosmological constant
problem" \cite{Weinberg:1988cp, Nobbenhuis:2004wn}, has not been solved yet.

A possible resolution to this problem is the anthropic selection of
the cosmological constant in the landscape of
vacua~\cite{Weinberg:1988cp, Susskind:2003kw, Nobbenhuis:2004wn}. In
the landscape of vacua, a vacuum can be unstable with respect to
tunneling to other vacua. The main interest of this paper is to
investigate how the stability of a vacuum is determined in the
non-linear massive gravity theory. We investigate how the tunneling
rate in general relativity (GR) is modified by the graviton mass
terms and clarify dependence on the model parameters. Moreover, an
interesting possibility will arise if the graviton mass depends on a
value of a scalar field \cite{Huang:2012, Saridakis:2012a,
Saridakis:2012jcap}, e.g., the tunneling field. In this case, the
effective cosmological constant of a vacuum can be larger than the
other while its potential energy is smaller because of the
contributions from the mass terms. It will be interesting to see how
the tunneling process proceeds in this case.

In this paper, we consider the Hawking-Moss (HM) solution
\cite{Hawking:1981fz} for a scalar field with minimal coupling to
gravity to understand the tunneling process in non-linear massive
gravity. We set up the model and found a bounce solution
corresponding to a HM solution. Based on this solution, we evaluate
the HM action and the contributions from the graviton mass terms. We
find they may either enhance or suppress the tunneling rate compared
to that in GR.

This paper is organized as follows.
In Sec. \ref{setup}, we setup the Lagrangian for our model.
In Sec. \ref{s:basic}, we formulate the equations of motion (EOM)
and solve the constraint equation.
In Sec. \ref{s:HM}, the HM solution is studied.
To analyze the effect of the graviton mass terms in details,
we consider several specific combinations of the parameters.

Throughout the paper, the Lorentzian metric signature is set to be $(-,+,+,+)$.

\section{Setup}\label{setup}

\begin{figure}
\includegraphics[height=6.5cm,keepaspectratio=true,angle=0]{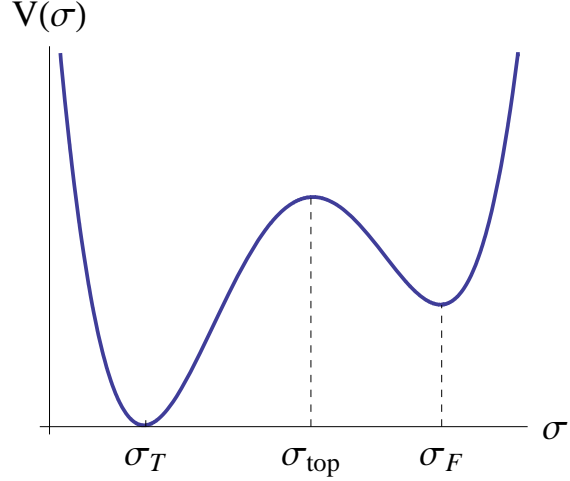}
\caption{Illustration of the potential $V(\sigma)$ with two local minima
$\sigma_F$ and $\sigma_T$, which correspond to the false and
true vacuum respectively. Here the $\sigma_{\rm top}$ labels
its local maximum value.}
 \label{fig:potential}
\end{figure}

We study a tunneling of a minimally coupled scalar field $\sigma$ between
two vacua in non-linear massive gravity.
It is described by a 4-dimensional metric $g_{\mu\nu}$, a fiducial
 metric $G_{ab}$, and the St\"{u}ckelberg fields $\phi^a$. \footnote{
We use Greek letters $\mu, \nu,...$ for the spacetime indices
and the Latin letters $i,j,...$ for the space indices,
while the Latin indices $a,b,...$ for the internal space (Lorentz frame)
indices. Repeated indices are understood to be summed over unless
otherwise stated.}
The action is given by \footnote{We use the natural units throughout this paper.}
\begin{align}\label{eq:action}
  S &= I_g+I_m, \\
  \label{eq:actiong}I_g &\equiv \int\mr{d}^4x~\sqrt{-g}\left[ \frac{R}{2}
 + m_g^2(\mc{L}_2 + \alpha_3\mc{L}_3 + \alpha_4\mc{L}_4)\right],
\\
  I_m &\equiv -\int\mr{d}^4x~\sqrt{-g}\left[ \frac{1}{2}(\pd \sigma)^2
 + V(\sigma) \right],
\end{align}
where
\begin{align}\label{eq:mass}
   {\cal L}_2 &= \frac{1}{2}\left(\left[{\cal K}\right]^2
-\left[{\cal K}^2\right]\right)\,,
\nonumber\\
   {\cal L}_3 &= \frac{1}{6}\left(\left[{\cal K}\right]^3
-3\left[{\cal K}\right]\left[{\cal K}^2\right]+2\left[{\cal K}^3\right]\right),
\nonumber\\
  {\cal L}_4 & = \frac{1}{24} \left(\left[{\cal K}\right]^4
-6\left[{\cal K}\right]^2\left[{\cal K}^2\right]
+3\left[{\cal K}^2\right]^2+8\left[{\cal K}\right]\left[{\cal K}^3\right]
-6\left[{\cal K}^4\right]\right),
\end{align}
and
\begin{align}\label{eq:k}
   {\cal K}^{\mu}_{\nu} \equiv \delta^{\mu}_{\nu}
 - \sqrt{g^{\mu\sigma}G_{ab}(\phi)\pd_{\nu}\phi^a\pd_{\sigma}\phi^b}.
    \end{align}
In this paper we assume $G_{ab}$ is non-dynamical as discussed below.
The potential $V(\sigma)$ is assumed to have two local minima $\sigma_F$
and $\sigma_T$, where the former corresponds to the
false vacuum, and a local maximum between them, $\sigma=\sigma_{\rm
top}$ (see Fig. \ref{fig:potential}).

Provided the action (\ref{eq:action}), the Euclidean action can be
obtained by $S_E=iS[x^0 \to ix_E^0]$.
Then, in the semiclassical limit, the tunneling rate per unit
time per unit volume is given in terms of the Euclidean action as
\begin{align}
    \Gamma/V= Ae^{-B}\,;
\quad
    B = S_E[\bar{g}_{\mu\nu,B},\bar{\phi}_B]
-S_E[\bar{g}_{\mu\nu,F},\bar{\phi}_F]\,,
\label{eq:rate}
\end{align}
where $\{ \bar{g}_{\mu\nu,B},\bar{\phi}_B \}$ is the so-called bounce solution,
or an instanton, a solution of the Euclidean equations of motion with
appropriate boundary conditions, and $\{ \bar{g}_{\mu\nu,F},\bar{\phi}_F \}$
is the solution staying at the false vacuum \cite{Coleman:1980}.
Among the bounce solutions, the one that gives the least action dominates
the tunneling.

Conventionally a bounce solution $\{ \bar{g}_{\mu\nu,B},\bar{\phi}_B \}$ is
 explored assuming $O(4)$-symmetry.
This is because it was proved that an $O(4)$-symmetric solution
gives the lowest action for a wide class of scalar-field
theories~\cite{Coleman:1977th}. It is therefore reasonable to assume the
same even in the presence of gravity \cite{Coleman:1980}. Hence the metric
would take the form,
 \begin{align}\label{eq:o4metric}
        g_{\mu\nu}\mr{d}x^{\mu}\mr{d}x^{\nu} = N(\xi)^2\mr{d}\xi^2
 + a(\xi)^2\Omega_{ij}\mr{d}x^i\mr{d}x^j,
    \end{align}
where the second term $\Omega_{ij}\mr{d}x^i\mr{d}x^j$ is
 the metric on a three-sphere $(K>0)$,
   \begin{align}\label{eq:3sphere}
           \Omega_{ij} &\equiv \delta_{ij}
 + \frac{K\delta_{il}\delta_{jm}x^l x^m}{1-K\delta_{lm}x^l x^m}\,.
    \end{align}
However, in massive gravity, this may not be always true if the
fiducial metric $G_{ab}$ does not respect the symmetry.

In this paper, to avoid possible complications due to the nature
of the fiducial metric, we assume that it is given
by the de Sitter metric with the ``Hubble parameter" $F$,
namely,
\begin{align}\label{eq:reference}
 G_{ab}(\phi)\mr{d}\phi^a\mr{d}\phi^b
&\equiv -(\mr{d}\phi^0)^2 +b(\phi^0)^2\Omega_{ij} \mr{d}\phi^i\mr{d}\phi^j,
\end{align}
where
\begin{align}\label{eq:bdefinition}
        b(\phi^0) &\equiv F^{-1}\sqrt{K}\cosh(F\phi^0)\,.
\end{align}
Then an $O(4)$-symmetric solution can be obtained by setting
\begin{align}\label{eq:o4scalars}
    \phi^0 = f(\xi)\,, \quad \phi^i = x^i\,.
\end{align}
The absence of the BD ghost has been proved for a general fiducial
metric in~\cite{Hassan:2011tf} and the de Sitter fiducial metric has
been recently investigated in~\cite{Langlois:2012hk}. In contrast to
the Minkowski fiducial metric, the de Sitter metric is shown to lead
to a wider range of cosmological
 solutions~\cite{Langlois:2012hk}.

At this level, the origin of the fiducial metric is unclear:
It may be a non-dynamical metric of the St\"{u}ckelberg field space
or one of the two dynamical metrics in bimetric
gravity~\cite{Hassan:2011zd, Volkov:2011an, Comelli:2011zm}.
In the later case, it may be natural to use the Euclidean signature also
for the fiducial metric, which can be accomplished by making the
replacement, $\phi^0 \to i\phi_{\rm E}^0~(f \to if_E)$, in
Eq.~(\ref{eq:o4scalars}). In this case, the $O(4)$-symmetric solutions
would exist also in the case that the fiducial metric is given by the Minkowski metric and the bounce solution could be found in a similar way by
setting $b(\phi^0) \equiv \sqrt{K}\phi^0$.

Here, we stick to the Lorentzian signature for the fiducial metric
because it is non-dynamical in our case. Nevertheless, thanks to the
assumption of the de Sitter fiducial metric, we may adopt the $O(4)$-ansatz.

\section{Basic equations} \label{s:basic}
Here we derive basic equations.
First we write down the Euclidean version of the action (\ref{eq:action}).
Inserting the $O(4)$-ansatz (\ref{eq:o4metric}) and (\ref{eq:o4scalars})
into the Euclidean version of (\ref{eq:k}), we find
\begin{align}
        \mc{K}^0_0 = 1-\frac{\sqrt{-\dot{f}^2}}{N}\,,
\quad \mc{K}^i_j = \left(1-\frac{b}{a}\right)\delta^i_j\,,
\quad \mc{K}^i_0=0, \quad \mc{K}^0_i=0\,,
\end{align}
where the dot means a derivative with respect to the radial
coordinate, $\dot{~} \equiv \mr{d}/\mr{d}\xi$.
This gives
\begin{align}
        [\mc{K}] =
\left(1-\frac{\sqrt{-\dot{f}^2}}{N}\right) + 3\left(1-\frac{b}{a}\right),
\\
        [\mc{K}^2] =
 \left(1-\frac{\sqrt{-\dot{f}^2}}{N}\right)^2 + 3\left(1-\frac{b}{a}\right)^2,
\\
        [\mc{K}^3] =
\left(1-\frac{\sqrt{-\dot{f}^2}}{N}\right)^3 + 3\left(1-\frac{b}{a}\right)^3,
\\
        [\mc{K}^4] =
\left(1-\frac{\sqrt{-\dot{f}^2}}{N}\right)^4 + 3\left(1-\frac{b}{a}\right)^4.
\end{align}
Then the gravity action is reduced to
\begin{align}\label{eq:euclidean}
        I_{gE} = \int\mr{d}^4x_E~\sqrt{\Omega}
\left[-3KNa - \frac{3\dot{a}^2 a}{N}
 - m_g^2\left(L_{2E}+\alpha_3L_{3E}+\alpha_4L_{4E}\right)\right],
\end{align}
where
    \begin{subequations}\label{eq:mass}
     \begin{align}
       \mathcal{L}_{2E} &= 3a(a-b)\left(2Na-\sqrt{-\dot{f}^2}a-Nb\right),\\
        \mathcal{L}_{3E} &= (a-b)^2\left(4Na-3\sqrt{-\dot{f}^2}a-Nb\right), \\
        \mathcal{L}_{4E} &= (a-b)^3\left(N-\sqrt{-\dot{f}^2}\right),
         \end{align}
     \end{subequations}
and the action for the tunneling field is reduced to
    \begin{align}\label{eq:ematter}
        I_{mE} = \int\mr{d}^4x~a^3
\sqrt{\Omega}\left[\frac{1}{2N}\dot{\sigma}^2 + NV(\sigma)\right],
    \end{align}
which coincides with the Euclidean version of the action (2.8) in
Ref. \cite{Gumrukcuoglu:2011open} if we replace $f$ by $b(f)$.

\subsection{Constraint equation from the St\"{u}ckelberg field}

The introduction of the St\"{u}ckelberg field gives rise to
an additional constrait equation.
Variation of the action (\ref{eq:euclidean}) with respect to $f$ gives
\begin{align}\label{eq:EOM1}
        (i\dot{a}+Nb_{,f})\left[\left(3-\frac{2b}{a}\right)
 + \alpha_3\left(1-\frac{b}{a}\right)\left(3-\frac{b}{a}\right)
 + \alpha_4\left(1-\frac{b}{a}\right)^2\right] = 0\,,
\end{align}
where
\begin{align}
        b_{,f} \equiv\frac{db}{df}= \sqrt{K}\sinh(Ff)\,.
\end{align}
Note that the derivation of the above equation does not depends
on the choice of a branch of $\sqrt{-\dot{f}^2}$.

Solving Eq.~(\ref{eq:EOM1}), we obtain two branches:
\begin{list}{}{}
\item[-] Branch I\\
    \begin{align}\label{eq:I}
        Nb_{,f} = -i\dot{a}\,.
    \end{align}
\item[-] Branch II\\
    \begin{align}
        \left(3-\frac{2b}{a}\right)
 + \alpha_3\left(1-\frac{b}{a}\right)\left(3-\frac{b}{a}\right)
 + \alpha_4\left(1-\frac{b}{a}\right)^2 = 0\,.
    \end{align}
The solution to this equation is given by
    \begin{align}\label{eq:II}
 b = X_{\pm}a\,,
\qquad
X_{\pm} \equiv
\frac{1+2\,\alpha_3+\alpha_4\pm\sqrt{1+\alpha_3+\alpha_3^2-\alpha_4}}
{\alpha_3+\alpha_4}\,.
    \end{align}
\end{list}
Hereafter, we call the choice $X_+$ as Branch II$_+$ and
$X_-$ as Branch II$_-$.
In Appendix \ref{s:branch1}, it is shown that Branch I is equivalent to
Branch II with a different set of parameters in the case of the HM solution.
Hence, we concentrate on the Branch II solutions hereafter.

\subsection{Euclidean equations of motion}
Variation of the action (\ref{eq:euclidean}) and (\ref{eq:ematter})
with respect to $N$ gives
    \begin{align}\label{eq:eom0}
        3\left(\frac{\dot{a}}{Na}\right)^2 - \frac{3K}{a^2}
 = \frac{1}{2N^2}\dot{\sigma}^2 - V(\sigma) - \rho_{g}\,,
    \end{align}
where
    \begin{align}
        \rho_{g} \equiv -m_g^2\left(1-\frac{b}{a}\right)
\left[3\left(2-\frac{b}{a}\right)
+ \alpha_3\left(1-\frac{b}{a}\right)\left(4-\frac{b}{a}\right)
 + \alpha_4\left(1-\frac{b}{a}\right)^2\right]\,.
    \end{align}
Inserting the solution (\ref{eq:II}) into the above,
$\rho_g$ reduces to a cosmological constant,
    \begin{align}\label{eq:lambda}
 \rho_g(b/a=X_\pm)=
 \Lambda_\pm \equiv -\frac{m_g^2}{\left(\alpha_3+\alpha_4\right)^2}
\left[\left(1+\alpha_3\right)\left(2+\alpha_3 +2\,\alpha_3^2-3\,\alpha_4\right)
 \pm 2\,\left(1+\alpha_3 +\alpha_3^2-\alpha_4\right)^{3/2}\right]\,.
    \end{align}
Variation of the action (\ref{eq:ematter}) with respect to $\sigma$ gives
\begin{align}
 \frac{1}{Na^3}\left(\frac{a^3\dot{\sigma}}{N}\right)^{\cdot}
 - V_{,\sigma} = 0\,.
\end{align}

Introducing the proper radial coordinate $\tau \equiv \int N{\rm d}\xi$,
the equations are rewritten as
\begin{align}
 &\frac{3}{a^2}\left(\diff{}{a}{\tau}\right)^2 - \frac{3K}{a^2}
=\frac{1}{2}\left(\diff{}{\sigma}{\tau}\right)^2 - V(\sigma) - \Lambda_{\pm}\,,
 \label{eq:eom}
 \\
 &\diff{2}{\sigma}{\tau} + 3\left(\diff{}{a}{\tau}\right)\diff{}{\sigma}{\tau}
 -V_{,\sigma}(\sigma) = 0\,.
 \label{eq:eom1}
\end{align}
As explained in the previous section, the tunneling rate (\ref{eq:rate})
is given by the Euclidean action evaluated for a solution of
Eqs.~(\ref{eq:eom}) and (\ref{eq:eom1}). In the following,
we construct a HM solution and then evaluate the action for it.

\section{Hawking-Moss solution}\label{s:HM}

\subsection{Evaluation of the tunneling rate}\label{s:HM solution}
A HM solution can be found by setting the tunneling field to the
local maximum value, $\sigma(\xi) = \sigma_{\rm top}$.
 Then the equation of motion (\ref{eq:eom1}) is trivially satisfied
and the Euclidean Friedmann equation (\ref{eq:eom}) reduces to
    \begin{align}\label{eq:eom2}
        \frac{3}{a^2}\left(\diff{}{a}{\tau}\right)^2 - \frac{3K}{a^2}
 = - V(\sigma_{\rm top}) - \Lambda_{\pm} \equiv -\Lambda_{\pm, \rm{eff}}\,.
     \end{align}
Setting the boundary condition as $a_{\rm HM}(H_{\rm HM}\tau=\pm\pi/2)=0$
and assuming $\Lambda_{\pm, \rm{eff}}>0$,
the HM solution is obtained as
\begin{align}\label{eq:aHMsol}
   a_{\rm{HM}}(\tau)
= H_{\rm HM}^{-1}\sqrt{K}\cos\left(H_{\rm HM}\tau\right)\,.
\end{align}
 Here, we have introduced the Hubble parameter of the physical metric by
\begin{eqnarray}\label{eq:hubblehm}
  H_{\rm HM} \equiv  \sqrt{\frac{\Lambda_{\pm,{\rm eff}}}{3}}
=  \sqrt{\frac{V(\sigma_{\rm top}) + \Lambda_{\pm}}{3}}\,.
\end{eqnarray}
Inserting Eq.~(\ref{eq:eom2}) into the Euclidian action given by
Eqs.~(\ref{eq:euclidean}) and (\ref{eq:ematter}),
and using $N^{-1}\dot f={\rm d}f/{\rm d}\tau$, the total
action can be expressed as
    \begin{align}\label{eq:actionHM}
        S_E[a_{\rm HM}, \sigma_{\rm top}] = \int {\rm d}^3x\sqrt{\Omega}
\int^{\pi/2H_{\rm HM}}_{-\pi/2H_{\rm HM}}{\rm d}\tau~ a_{HM}^3
\left(2\Lambda_{\pm,\rm{eff}}-\frac{6K}{a_{\rm HM}^2}
 + m_g^2Y_\pm \sqrt{-\left(\diff{}{f_{\rm HM}}{\tau}\right)^2}\right)\,,
    \end{align}
where, for brevity, we have introduced the parameter $Y_\pm$
in terms of $X_{\pm}$ as
\begin{eqnarray}\label{eq:Y}
    Y_\pm\equiv3(1-X_{\pm})+3\alpha_3(1-X_{\pm})^2+\alpha_4(1-X_{\pm})^3\,.
\end{eqnarray}

We also need the solution for the St\"{u}ckelberg field $f$ to evaluate
the action. It is given in terms of $a_{\rm HM}$ as in Eq. (\ref{eq:II}),
\begin{align}\label{eq:bf}
        b_{\rm HM} = F^{-1}\sqrt{K}\cosh(Ff_{\rm HM}) = X_{\pm}a_{\rm{HM}}\,.
\end{align}
This can be solved for $f$ as
\begin{eqnarray}\label{eq:fHMsol}
      f_{\rm HM}(\tau)
=\frac{1}{F}\ln\left[\alpha_{\rm HM}\cos(H_{\rm HM}\tau)
\pm\sqrt{\alpha_{\rm HM}^2\cos^2(H_{\rm HM}\tau)-1}\right]\,,
\end{eqnarray}
where the parameter $\alpha$ is defined as
\begin{eqnarray}\label{eq:reparameter}
        \alpha_{\rm HM} \equiv  X_\pm\frac{F}{H_{\rm HM}}\,.
\end{eqnarray}
As clear from the above definition, the parameter $\alpha_{\rm HM}$ represents
the ratio of the Hubble parameter of the fiducial metric
and that of the physical metric.
Since $X_{\pm}$ should be positive for the constraint (\ref{eq:II})
to be satisfied, $\alpha_{\rm HM}$ is also positive.

Finally, we evaluate the derivative of $f$
since the action (\ref{eq:actionHM}) depends on the
 St\"{u}ckelberg field only through ${\rm d}f/{\rm d}\tau$.
Taking a derivative of Eq.~(\ref{eq:bf}) with respect to $\tau$, one obtains
\begin{eqnarray}\label{eq:fHM2der}
        \diff{}{f_{\rm HM}}{\tau}
=-\frac{X_{\pm}\sin(H_{\rm HM}\tau)}{\sinh(Ff_{HM}(\tau))}\,,
\end{eqnarray}
where $\sinh(Ff_{HM})$ can be calculated from Eq.~(\ref{eq:bf}) as
\begin{align}\label{eq:sinhf}
        \sinh(Ff_{HM}) = \pm\sqrt{\alpha_{\rm HM}^2\cos^2(H_{\rm HM}\tau)-1}\,.
\end{align}
Though we have obtained ``$\pm$" branches for ${\rm d}f/{\rm d}\tau$ here,
the sign is irrelevant because
the action (\ref{eq:actionHM}) depends only on its square,
\begin{align}\label{eq:sqfHM2der}
\left( \diff{}{f_{\rm HM}}{\tau}\right)^2
= \frac{X_{\pm}^2\sin^2(H_{\rm HM}\tau)}{\alpha_{\rm HM}^2\cos^2(H_{\rm HM}\tau)-1}\,.
\end{align}

Here we note that we can evaluate ${\rm d}f_E/{\rm d}\tau~(f_E=-if)$ in a similar way
for the Minkowski fiducial metric by setting $b_{\rm HM}=\sqrt{K}f_E$ in
Eq.~(\ref{eq:bf}). Taking a derivative with respect to $\tau$, one obtains
\begin{align}
 \diff{}{f_{E, {\rm HM}}}{\tau}
=-X_{\pm}\sin(H_{\rm HM}\tau)\,.
\end{align}
The square of this equation coincides with Eq.~(\ref{eq:sqfHM2der})
with $\alpha_{\rm HM} = 0$ after making the replacement $f_E=-if$.
Hence, the analysis below includes the results for the Minkowski
fiducial metric.


Note that the solution for $\phi^0_{\rm HM}=f_{\rm HM}$ does not diverge in the limit $m_g \to 0$.
Hence, our results continuously reduce to the results in GR.
This is consistent with the observations that there is no vDVZ discontinuity in the self-accelerating branch \cite{Koyama:2012prda, Koyama:2011jhep, Gumrukcuoglu:2011perturb}.

It should also be noted that the induced metric
$G_{ab}\pd_{\mu}\phi^a\pd_{\nu}\phi^b$ in the case of $\alpha_{\rm
HM}>1$ has a ``singularity", where $\dot{f}$ becomes infinite, and
becomes neither pure imaginary nor pure real. The singularity would
also appear even if we had started from the Euclidean signature for
the fiducial metric. In this case, the four sphere for the fiducial
metric would crunch before the physical one does. For this reason,
we concentrate on the case where $\alpha_{\rm HM} \leq 1$.

As easily seen from the solution~(\ref{eq:fHMsol}), the
St\"{u}ckelberg field becomes complex in general.
However, as we will see,  the action turns out to be real for
$\alpha_{\rm HM} \leq 1$.
We also note that if we start from the Euclidean fiducial metric,
$f_{\rm HM}$ and ${\rm d}f_{\rm HM}/{\rm d}\tau$ become real
for $\alpha_{\rm HM} \leq 1$.

\subsubsection{Action for the case $\alpha\leq1$}
In the action (\ref{eq:euclidean}), the only non-trivial term is the one
that depends on ${\rm d}f/{\rm d}\tau$. Hence, we first evaluate this part.
 Substituting the solution (\ref{eq:sqfHM2der}), we find
    \begin{eqnarray}
 m_g^2Y_\pm\int_{-\pi/2H_{\rm HM}}^{\pi/2H_{\rm HM}}
{\rm d}\tau~ a_{HM}^3\sqrt{-\left(\diff{}{f_{\rm HM}}{\tau}\right)^2}
&=& \frac{2m_g^2 Y_{\pm} X_{\pm}}{H_{\rm HM}^4}K^{\frac{3}{2}}
\int_{0}^{\pi/2}{\rm d}(H_{\rm HM}\tau)~\frac{\cos^3(H_{\rm HM}\tau)
\sin(H_{\rm HM}\tau)}{\sqrt{1-\alpha_{\rm HM}^2\cos^2(H_{\rm HM}\tau)}}
\nonumber\\
&=& \frac{2m_g^2 Y_{\pm}X_\pm}{3\alpha_{\rm HM}^4H_{\rm HM}^4}K^{\frac{3}{2}}
\sqrt{1-z^2}(2+z^2)|_{\alpha_{\rm HM}}^0, \quad (z\equiv\alpha_{\rm HM}\cos(H_{\rm HM}\tau))
 \nonumber\\
&=& \frac{2m_g^2 Y_{\pm}X_\pm}{3\alpha_{\rm HM}^4H_{\rm HM}^4}K^{\frac{3}{2}}
\left[2-\sqrt{1-\alpha_{\rm HM}^2}(2+\alpha_{\rm HM}^2)\right] \,.
\end{eqnarray}
As claimed, the action is real.

\subsubsection{Hawking-Moss action}
From the result of the previous subsection,
we obtain the Euclidean action as
\begin{align}\label{eq:actiona}
 S_{\rm HM}&\equiv
 S_E[a_{\rm HM},\sigma_{\rm top}]
= -\frac{8\pi^2}{H_{\rm HM}^2}\left[1 - \frac{Y_{\pm}X_\pm}{6\alpha_{\rm HM}^4}
\left(\frac{m_g}{H_{\rm HM}}\right)^2
\left(2-\sqrt{1-\alpha_{\rm HM}^2}(2+\alpha_{\rm HM}^2)\right)\right]
\nonumber \\
&
= -\frac{8\pi^2}{H_{\rm HM}^2}\left[1 - \frac{Y_{\pm}X_\pm}{6}
\left(\frac{m_g}{H_{\rm HM}}\right)^2A(\alpha_{\rm HM})\right];
\quad A(\alpha)\equiv
\frac{2-\sqrt{1-\alpha^2}(2+\alpha^2)}{\alpha^4}\,,
\end{align}
where $\int d^3x\sqrt{\Omega}=2\pi^2K^{-3/2}$ has been used.
As previously explained, we consider only the case $\alpha_{HM} \leq1$
 since a singularity would appear otherwise.
In addition to the standard first term determined by the
Hubble parameter $H_{\rm HM}$ (which nevertheless contains
a contribution of the mass term $\Lambda_{\pm}$),
there apprears a mass-dependent term.
The presence of this second term may be regarded as
the genuine effect of the graviton mass.
We also note here that the function $A(\alpha)$ is regular at $\alpha=0$, which corresponds to the case of the Minkowski fiducial metric.

As given by Eq.~(\ref{eq:rate}), the tunneling rate is determined
by the difference between the HM action $S_{\rm HM}$
given by Eq.~(\ref{eq:actiona}) and
the false vacuum action $S_{\rm F}=S_E(\bar g_{\mu\nu,F},\bar\phi_F)$.
The latter is given similarly as Eq. (\ref{eq:actiona})
by replacing the Hubble parameter by
\begin{align}\label{eq:hubblef}
        H_{\rm F} \equiv \sqrt{\frac{V(\sigma_F)+\Lambda_{\pm}}{3}}\,,
\end{align}
and $\alpha$ by its false vacuum value,
$\alpha_{\rm F}\equiv X_{\pm}F/H_{\rm F}$.
We note that $\alpha_{\rm F}$ should not exceed unity either.
Since the potential energy at the false vacuum is always
smaller that at the maximum, $V(\sigma_F)<V(\sigma_{\rm top})$,
we have
\begin{eqnarray}
\alpha_{\rm F}>\alpha_{\rm HM}\equiv \frac{X_{\pm}F}{H_{\rm HM}}\,.
\end{eqnarray}
Hence the constraint $\alpha_{\rm F}\leq1$ gives a tighter constraint on
the potential energy than that from $\alpha_{\rm HM} \leq 1$.
Conversely, if the false vacuum exists without a singularity
in the St\"{u}ckelberg field, a regular HM solution always exists.

To see the effect of the second term, let us compare the
current result with the HM action in GR with the same value of the Hubble
parameter,
$B^{\rm (GR)}\equiv 8\pi^2(-H_{\rm HM}^{-2}+H_{\rm F}^{-2})$.
We have
\begin{eqnarray}
\Delta B\equiv
B-B^{\rm (GR)}=\frac{4\pi^2}{3}Y_{\pm}X_{\pm}\left(\frac{A(\alpha_{\rm HM})}{H_{\rm HM}^4}
-\frac{A(\alpha_{\rm F})}{H_{\rm F}^4}\right)m_g^2\,.
\end{eqnarray}
Note that the function $A(\alpha)$ is positive
and a monotonically increasing function of $\alpha$ for $0<\alpha\leq1$.
Hence $0<A(\alpha_{\rm HM})<A(\alpha_{\rm F})$.
Since $H_{\rm HM}^{-1}<H_{\rm F}^{-1}$, this means the sign of
$\Delta B$ is determined by the sign of $Y_{\pm}$.
Namely, for $Y_{\pm}>0$ ($<0$), we have $\Delta B<0$ ($>0$),
and hence the tunneling rate
is enhanced (suppressed) relative to the case of GR.

\subsection{Parameter dependence of the tunneling rate}
In the above, we have seen the effect of the graviton mass on
the tunneling rate is determined by the sign of $Y_\pm$.
In this subsetion, we analyze the dependence of the tunneling rate on
the parameters $\alpha_3$ and $\alpha_4$ by
examining the sign of $Y_{\pm}$.

\subsubsection{General case with $m_g \neq 0$}
\label{mgneqzero}

\begin{figure}
\includegraphics[height=6.5cm,keepaspectratio=true,angle=0]{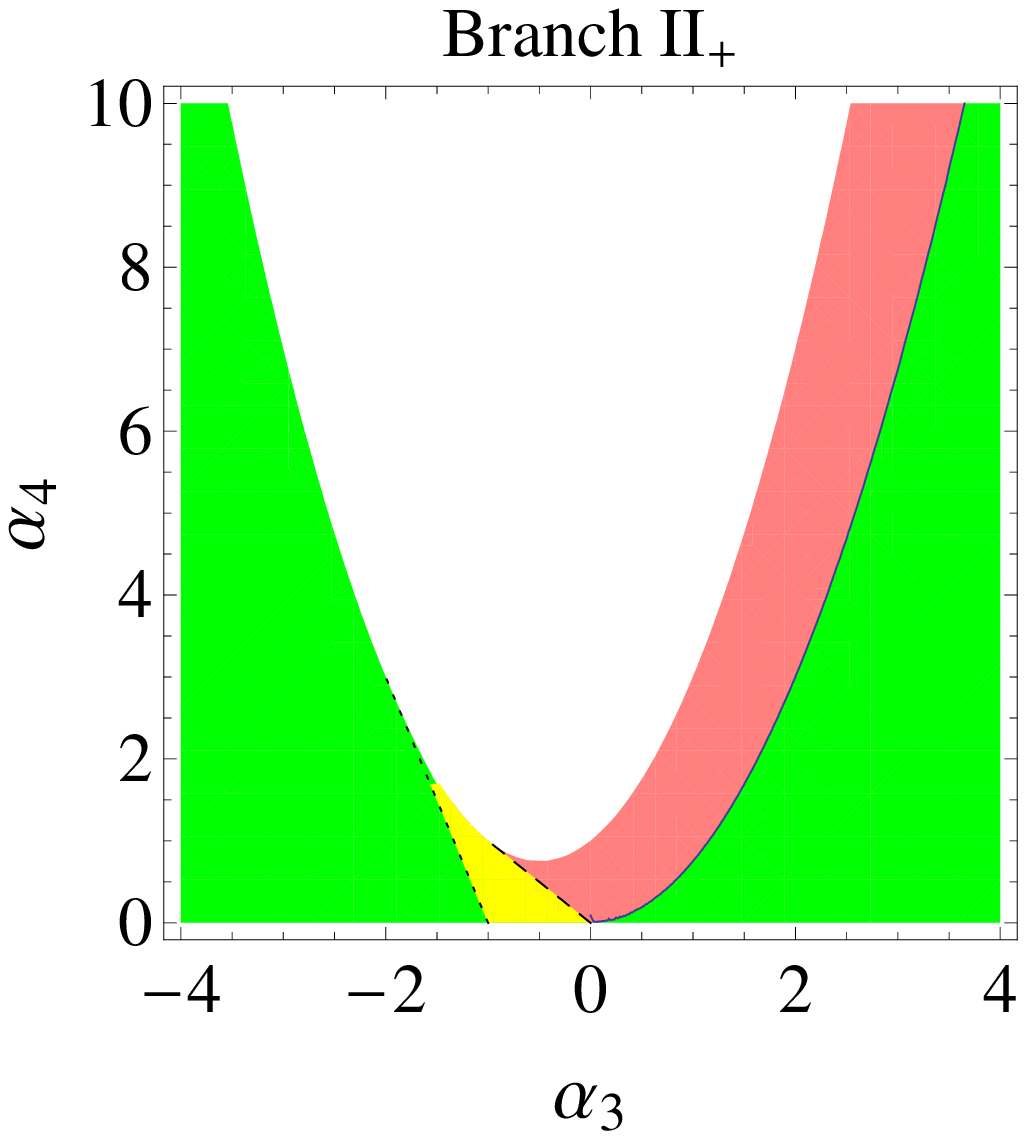}
\includegraphics[height=6.5cm,keepaspectratio=true,angle=0]{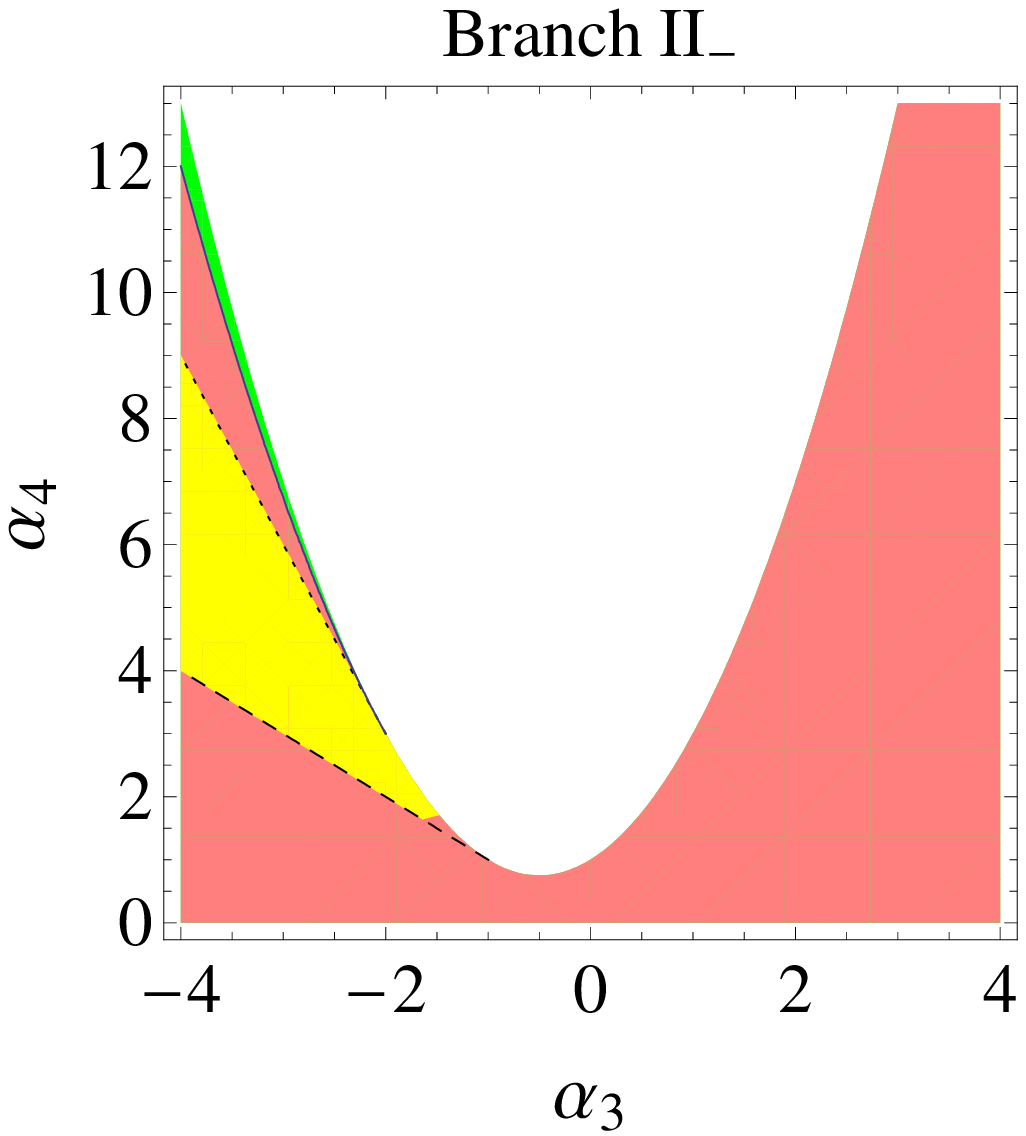}
\caption{The sign of $Y_\pm$ is shown for Branch II$_{+}$ (left panel) and
Branch II$_{-}$ (right panel) solutions. The white and yellow regions
corresponds to $1+\alpha_3+\alpha_3^2-\alpha_4<0$ and $X_\pm<0$
respectively, which should be excluded since the cosmological
solution does not exist in these
regions~\cite{Gumrukcuoglu:2011open, Gumrukcuoglu:2011perturb}.
The green regions correspond to $Y_\pm>0$, which implies that the HM
tunneling rate is enhanced in comparison to the GR case for the
same Hubble parameter $H_{\rm HM}$, while the pink regions
correspond to $Y_\pm<0$, which implies that the HM tunneling
rate is suppressed. Along the solid lines (defining the boundary between
green and pink regions), $X_\pm=1$ and hence $Y_\pm=0$. Hence the HM
solution (\ref{eq:actiona}) reduces to that in GR. The dotted
lines denote $X_\pm=0$ where the solution ceases to exist.
Along the dashed lines, $X_\pm$ diverges and thus defines
another boundary of the solution space.}
\label{fig:Xpm}
\end{figure}

As shown in the previous subsection, the change in the tunneling rate
is determined by the sign of $Y_{\pm}$, which is given in terms of
$X_{\pm}$ as given by Eq.~(\ref{eq:Y}).
Let us recapitulate it here:
\begin{align}
X_{\pm} &=
\frac{1+2\,\alpha_3+\alpha_4\pm\sqrt{1+\alpha_3+\alpha_3^2-\alpha_4}}
{\alpha_3+\alpha_4}\,,
\nonumber\\
Y_\pm &=3(1-X_{\pm})+3\alpha_3(1-X_{\pm})^2+\alpha_4(1-X_{\pm})^3\,.
\label{XYpm}
\end{align}
In Fig.~\ref{fig:Xpm}, we show the sign of $Y_\pm$
in the parameter space $(\alpha_3,\alpha_4)$
for Branch II$_{+}$ (left panel) and Branch II$_{-}$ (right panel) solutions.
As seen from the figure, the tunneling rate is enhanced (suppressed) for
Branch II$_+$ (Branch II$_-$) in a large region of the parameter space.
It should be noted again that $\alpha_F\leq1$ gives rise to
a constraint on $V(\sigma_{\rm{F}})$.

 Next, we focus on some limiting cases and
analytically investigate the parameter dependence of the
tunneling rate.

\subsubsection{$\alpha_3=0$}

In this case, from Eq.~(\ref{XYpm})
\begin{eqnarray}\label{eq:a30}
X_\pm=\frac{1}{\alpha_4}\left(1+\alpha_4\pm\sqrt{1-\alpha_4}\right)
\,,
\quad
Y_\pm=(1-X_\pm)\left(2+\alpha_4(1-X_\pm)^2\right)\,,
\end{eqnarray}
and from Eq.~(\ref{eq:lambda}),
\begin{eqnarray}
\qquad\Lambda_\pm=-\frac{m_g^2}{\alpha_4^2}
\left[2-3\alpha_4\pm2(1-\alpha_4)^{\frac{3}{2}}\right].
\end{eqnarray}
Note that the parameter $\alpha_4$ should satisfy $\alpha_4\leq1$
for $X_\pm$ to be real. In the following, we evaluate the HM action
 for Branch II$_+$ and Branch II$_{-}$ separately.
\\

\noindent $\bullet$ Branch II$_+$
\\

In this branch, we have
\begin{eqnarray}\label{eq:a30plus}
X_+=\frac{1}{\alpha_4}\left(1+\alpha_4+\sqrt{1-\alpha_4}\right)\,,
\quad
Y_+ = -\frac{2}{\alpha_4^2}(1+\sqrt{1-\alpha_4})
(1+\alpha_4 + \sqrt{1-\alpha_4})\,,
\end{eqnarray}
and
\begin{align}
\Lambda_+=-\frac{m_g^2}{\alpha_4^2}
\left[2-3\alpha_4+2(1-\alpha_4)^{\frac{3}{2}}\right]\,.
\end{align}

From the expression above, one immediately finds that $\alpha_4=0$
can be a singular point for Branch II$_+$. To study this case, we set
$\alpha_3=\alpha_4=\epsilon\ll1$ and take the limit $\epsilon\to0$.
Then one obtains
\begin{eqnarray}
\Lambda_+(\alpha_3=\alpha_4=0)=-\frac{m_g^2}{2\epsilon^2}
\left[2+\frac{3\epsilon^2}{2}+\mathcal {O}(\epsilon^3)\right]\,,
\end{eqnarray}
which leads to
\begin{equation} \label{eq:2}
\lim_{\epsilon\rightarrow0}\Lambda_+(\alpha_3=\alpha_4=0)
=-\infty\,.
\end{equation}
So $\alpha_3=\alpha_4=0$ is singular in this branch, which should be excluded
in the analysis. From the condition $\alpha_{\rm{F}}\leq1$, we
obtain the constraint on $V(\sigma_{\rm{F}})$ by using
Eqs.~(\ref{eq:reparameter}) and (\ref{eq:a30plus}):
\begin{eqnarray}\label{eq:mgplus1}
V(\sigma_{\rm{F}}) \geq \frac{\left(2\gamma_4^{3} + 3\gamma_4^2-1\right)m_g^2
+3\left(\gamma_4^2 - \gamma_4 -2 \right)^2F^2}{(1-\gamma_4^2)^2}
=\frac{(2\gamma_4-1)m_g^2+3(\gamma_4-2)^2F^2}{(\gamma_4-1)^2}\,,
\end{eqnarray}
where
\begin{eqnarray}
\gamma_4 \equiv \sqrt{1-\alpha_4}\ (\geq0)\,.
\end{eqnarray}
An interesting consequence of this constraint is the absence of a
(meta-)stable vacuum state below this bound. For example, even if
there is a potential minimum below this bound, tunneling to such a
state may be prohibited.

In terms of $\gamma_4$, Eqs.~(\ref{eq:a30plus}) can be re-expressed as
\begin{align}
X_+=\frac{\gamma_4-2}{\gamma_4-1}\,,
\quad
Y_{+}= 2\frac{\gamma_4-2}{(\gamma_4-1)^2}\,.
\end{align}
Thus the condition $X_+>0$ requires either $\gamma_4>2$ ($\alpha_4<-3$)
for which we have $Y_+>0$, or $0\leq\gamma_4<1$ ($1\geq\alpha_4>0$)
for which $Y_+<0$. Therefore the tunneling rate is enhanced for
$\alpha_4<-3$ and suppressed for $0<\alpha_4 \leq 1$.
The range $-3<\alpha_4<0$ is excluded because a solution ceases
to exist there.

In order for the solution to be a four-sphere, the effective
cosmological constant must be positive definite, $\Lambda_{+,{\rm eff}}>0$.
We find this is always satisfied, since
\begin{eqnarray}
\Lambda_{+,{\rm eff}}\geq V(\sigma_{\rm{F}})+\Lambda_+
\geq 3F^2X_\pm^2>0\,.
\end{eqnarray}
We again note that $\gamma_4=1~(\alpha_4=0)$ is a singular point,
which coincides with the conclusion that Branch II$_+$ does not
exist for $\alpha_3=\alpha_4=0$.
\\

\noindent
$\bullet$ Branch II$_-$
\\

In this branch, we have
\begin{eqnarray}\label{eq:a30minus}
&&X_-=\frac{1}{\alpha_4}\left(1+\alpha_4-\sqrt{1-\alpha_4}\right)
=\frac{\gamma_4+2}{\gamma_4+1}\,,
\nonumber\\
&&Y_- = \frac{2}{\alpha_4^2}(-2+2\sqrt{1-\alpha_4}+\alpha_4\sqrt{1-\alpha_4})
=-2\frac{\gamma_4+2}{(\gamma_4+1)^2}\,,
\end{eqnarray}
and
\begin{align}
\Lambda_-=-\frac{m_g^2}{\alpha_4^2}
\left[2-3\alpha_4-2(1-\alpha_4)^{\frac{3}{2}}\right]
=\frac{2\gamma_4+1}{(\gamma_4+1)^2}m_g^2\,,
\end{align}
where, as before, $\gamma_4=\sqrt{1-\alpha_4}$ and the parameter
$\alpha_4$ should not exceed unity. We note that in contrast with
Branch II$_+$, here $\alpha_3=\alpha_4=0$ is regular, which implies
a finite $\Lambda_+(\alpha_3=\alpha_4=0)=3m_g^2/4$.

Similar to the case of Branch II$_+$, we find a
constraint on $V(\sigma_F)$ on Branch II$_-$ from the condition
$\alpha_4\leq1$ as
\begin{eqnarray}\label{eq:constrV2-}
 V(\sigma_F)\geq\frac{\left(-2\gamma_4^{3}
+ 3\gamma_4^2-1\right)m_g^2
+3(\gamma_4^2 + \gamma_4 -2)^2F^2}{(1-\gamma_4^2)^2}
=\frac{-(2\gamma_4+1)m_g^2+3(\gamma_4+2)^2F^2}{(\gamma_4+1)^2}\,.
\end{eqnarray}
It is clear that the limit $\gamma_4\to1~(\alpha_4\to0)$ is regular
in this case of Branch II$_-$, which is consistent with the case
$\alpha_3=\alpha_4=0$ studied previously. We note that
there is no restriction on the range of $\alpha_4$
except for the condition that $\alpha_4\leq1$.

From Eqs.~(\ref{eq:a30minus}), we see that $Y_-$ is always
negative. Therefore, the tunneling rate is always suppressed in this case.

\subsubsection{Vanishing $\mathcal {L}_2$ term}

In this subsection, we consider a special case when $\mathcal{L}_2$
vanishes but $\mathcal {L}_3$ and $\mathcal {L}_4$ are non-vanishing.
To achieve this, we take the limit $m_g^2 \to 0$,
while keeping $A_3\equiv m_g^2\alpha_3$ and $A_4\equiv m_g^2\alpha_4$
finite. Substituting $A_3$ and $A_4$ for $\alpha_3$ and $\alpha_4$ in
Eq.~(\ref{eq:lambda}) and taking $m_g^2\rightarrow0$,
$\Lambda_\pm$ is calculated as
\begin{equation} \label{eq:lambdaa20Lam}
  \Lambda_\pm=-\frac{2A_3^3}{(A_3+A_4)^2}\left[1\pm\rm{sgn}(A_3)\right]\,,
\end{equation}
where $\rm{sgn}(A_3)\equiv|A_3|/A_3$ denotes the sign of the parameter $A_3$.
The sign function appears due to the presence of the
square root in Eq.~(\ref{eq:lambda}).
Similarly, from Eq.~(\ref{eq:II}), $X_\pm$ can be obtained as
\begin{eqnarray}\label{eq:lambdaa20X}
        X_\pm=1+\frac{A_3}{A_3+A_4}\left[1\pm\rm{sgn}(A_3)\right]\,.
\end{eqnarray}
Without loss of generality, we assume $A_3>0$
since the flip of sign only exchanges the roles of Branch II$_+$
and Branch II$_-$.
\\

\noindent
$\bullet$ Branch II$_+$
\\

In this branch, from Eqs.~(\ref{eq:lambdaa20Lam}) and (\ref{eq:lambdaa20X}),
 we have
\begin{equation} \label{eq:L20+1}
\Lambda_+=-\frac{4A_3^3}{(A_3+A_4)^2}<0\,,\qquad X_+
=1+\frac{2A_3}{A_3+A_4}\,,
\end{equation}
and
\begin{align}
 m_g^2Y_+ &= 3A_3(1-X_+)^2 + A_4(1-X_+)^3 \nonumber \\
        &= (1-X_+)^2X_+A_3\,.
\end{align}
Thus $m_g^2Y_-$ is always positive provided $X_->0$.
From Eq.~(\ref{eq:L20+1}), this is guaranteed if
the parameter $A_4$ is either in the range $A_4<-3A_3$ or $-A_3<A_4$.
The tunneling rate is always enhanced in this case.

Next let us consider the condition $\alpha_{\rm F}<1$.
From Eq.~(\ref{eq:L20+1}), we find
\begin{equation} \label{eq:L20+al}
\alpha^2_{\rm F}
=\frac{3F^2(3A_3+A_4)^2}{(A_3+A_4)^2V(\sigma_{{\rm F}})-4A_3^3}\leq1\,.
\end{equation}
Hence the constraint on $V(\sigma_{{\rm F}})$ is found as
\begin{equation}\label{eq:constrV1+}
V(\sigma_{\rm{F}})\geq\frac{3F^2(3A_3+A_4)^2+4A_3^3}{(A_3+A_4)^2}\,.
\end{equation}
Again the positivity of the effective cosmological
constant, $\Lambda_{+, \rm{eff}}>0$, is guaranteed because
\begin{eqnarray}
\Lambda_{+,{\rm eff}}\geq V(\sigma_{\rm{F}})+\Lambda_+
\geq 3F^2X_+^2>0\,.
\end{eqnarray}
\\

\noindent
$\bullet$ Branch II$_-$
\\

In this branch, from Eqs.~(\ref{eq:lambdaa20Lam}) and (\ref{eq:lambdaa20X}),
we find
\begin{equation} \label{eq:L20+}
\Lambda_-=0\,,\qquad X_- =1\,, \qquad m_g^2Y_{-}=0\,.
\end{equation}
This implies that the action in this case reduces to that of GR.
There is no difference from GR in this branch.

From Eq.~(\ref{eq:L20+}), the condition $\alpha_{\rm F}\leq1$
is expressed as
\begin{equation} \label{eq:L20+al}
\alpha_{\rm{F}}^2=\frac{3F^2}{V(\sigma_{\rm{F}})}\leq1\,,
\end{equation}
which leads to the constraint on $V(\sigma_{\rm{F}})$ as
\begin{equation}\label{eq:constrV1-}
V(\sigma_{\rm{F}})\geq3F^2\,.
\end{equation}

\section{Summary and discussion}


In this paper, we investigated the stability of a vacuum in
the landscape of vacua in a theory of non-linear massive
gravity. For this purpose, we derived the Hawking-Moss
(HM) solution in a simple case where a tunneling scalar field
is minimally coupled with gravity and evaluated its Euclidean action.
 We obtained three branches of the solution, which we labeled
as Branch I and Branch II$_{\pm}$, where Branch II$_{\pm}$ corresponds
to the self-accelerating branch
in Ref.~\cite{Gumrukcuoglu:2011open, Gumrukcuoglu:2011perturb}.
We analyzed the contribution of the graviton mass terms
to the Euclidean action, hence to the tunneling rate.
We focused on Branch-II$_{\pm}$ HM solutions because
a Branch-I HM solution is shown to be equivalent to a Branch-II
solution with a different set of the model parameters.

The Euclidean action of the HM instanton is found to have two distinct
terms. In addition to the standard contribution determined by the Hubble
parameter of the HM solution, we obtained a mass-dependent
non-standard term. To study the effect of this latter
non-standard contributions, we compared the tunneling rate to that
in GR for the same value of the Hubble parameter.

For Branch II$_\pm$, the enhancement or suppression of the
tunneling rate relative to GR is determined by the sign of
a quantity, denoted by $Y_{\pm}$, and it depends on the
 model parameters $\alpha_3$ and $\alpha_4$.
We found that in a wide area of the parameter space
the tunneling rate is enhanced for Branch II$_{+}$
and suppressed for Branch II$_{-}$.

We also found that the solution can exist only if the
Hubble parameter of the ratio of the physical metric
to the fiducial one is greater than $X_\pm$. As a result, the form of the potential
is constrained. This seems to imply that tunneling from or to a vacuum
whose energy density is less than a critical value is prohibited.
This result is in sharp contrast to GR for which
there is no bound on the minimum value of the potential.

As tunneling mediated by a HM instanton is considered to be
equivalent to stochastic process of going over the barrier due to
large vacuum fluctuations intrinsic to light scalar fields in de
Sitter space~\cite{Starobinsky:1986fx}, it is interesting to
understand the role of the graviton mass terms in the context of
stochastic dynamics.

We found that the tunneling rate can be enhanced or suppressed
depending on the model parameters for a simple case of a tunneling
field minimally couples with gravity. Though we discussed this
simple case as a first step, it will be interesting to see how the
tunneling process proceeds when the model parameters such as the
graviton mass depend on the tunneling field. In this case, the
effective cosmological constant of a vacuum can be larger than the
other while its potential energy is smaller because of the tunneling
field dependence of the graviton mass.


Finally, let us comment on the limitation of our analysis.
Since massive gravity exhibits the strong coupling at high energy scales,
it requires UV completion above its cutoff scale.
Namely, our analysis cannot apply to an arbitrarily small size instanton
nor to the potential of an arbitrarily high energy scale,
though the limitation depends on how massive gravity is UV completed.
The cutoff scale of massive gravity is determined by the graviton mass and
it is very low if the graviton mass is of the order of the present Hubble
parameter to explain the current accelerated expansion of the universe.
However, as mentioned above, the graviton mass may depend on the tunneling
field and may have a larger value in the early universe.
In this case, our analysis can be applied to the tunneling process
at high energy scales.

\appendix

\section{HM solution for Branch I}\label{s:branch1}

In the text, we focused on HM solutions based on
Branch II constraint (\ref{eq:EOM1}).
Here, we show that the HM solution in Branch I
is equivalent to the one in Branch II with a different set
of the model parameters.

Inserting Eq.~(\ref{eq:bdefinition}) into Eq.~(\ref{eq:I}),
we obtain
\begin{eqnarray}
F^2b^2+a^{\prime2}=K\,.
\end{eqnarray}
Hence
\begin{eqnarray}
b=\frac{\sqrt{K-a^{\prime2}}}{F}\,.
\end{eqnarray}
Inserting this into Eq.~(\ref{eq:eom0}) and setting
$\sigma \equiv \sigma_{\rm top}$, we obtain an algebraic equation,
\begin{align}
     3F^2\theta^2 - V(\sigma_{\rm top}) - \rho_g= 0\,,
\end{align}
where
\begin{align}
   \rho_{g} \equiv
  -m_g^2\left(1-\theta\right)\left[3\left(2-\theta\right)
 + \alpha_3\left(1-\theta\right)\left(4-\theta\right)
 + \alpha_4\left(1-\theta\right)^2\right]\,,
\quad
\theta\equiv\frac{\sqrt{K-a^{\prime2}}}{Fa}=\frac{b}{a}\,.
\end{align}
Hence, $\theta=b/a$ becomes constant. The HM solution in Branch I is
 obtained as
\begin{eqnarray}
a_{\rm HM}^I(\tau)
=(\theta F)^{-1}\sqrt{K}\cos\left(\theta F\tau\right)\,.
\end{eqnarray}
Comparing this with Eqs.~(\ref{eq:II}) and (\ref{eq:aHMsol}),
we see that the form of the solutions are the same except
that $X_{\pm}$ and $H_{\rm HM}$ are replaced by $\theta$ and
$\theta F$, respectively. We note that the quantity corresponding to
$\alpha \equiv X_{\pm}F/H_{\rm HM}$ in Branch I is unity.

Thus, we conclude that the analysis of the HM solutions
for Branch II can be directly applied to those for Branch I
if we make the replacement $X_{\pm} \to \theta$ and
$H_{\rm HM}\rightarrow\theta F$.
Note that $\theta$ depends on $F$, $m_g$, and $V(\sigma_{\rm top})$,
as well as $\alpha_3$ and $\alpha_4$. Hence, the dependence on the
model parameters becomes more complicated in Branch I than in Branch II.

\section{List of symbols}
Throughout the paper, there appear many parameters which are denoted
by various symbols.
For convenience, we list some of them who play important roles
in the table \ref{tab:symbols}.
\begin{table}[!h]
\tabcolsep 5mm
\caption{List of symbols}
\label{tab:symbols}
\centering
\begin{tabular}{r@{}lr@{}lr@{}l}
\hline
\multicolumn{2}{c}{Symbols}&\multicolumn{2}{c}{Definition}
&\multicolumn{2}{c}{First
appearance}
\\ \hline
$X_\pm$&   &See Eq. (\ref{eq:II})& &(\ref{eq:II}) \\
$\Lambda_\pm$&      &See Eq. (\ref{eq:lambda})&  &(\ref{eq:lambda}) \\
$\Lambda_{\pm, \rm{eff}}$&  &$V(\sigma_{\rm top})+\Lambda_{\pm}$& &(\ref{eq:eom2})\\
$Y_\pm$&      &See Eq. (\ref{eq:Y})&  &(\ref{eq:Y})\\
$H_{\rm HM}$&      &$\sqrt{\Lambda_{\pm, \rm eff}/3}$&   &(\ref{eq:hubblehm})\\
$\alpha_{\rm HM}$&      &$FX_\pm/H_{\rm HM}$&  &(\ref{eq:reparameter})\\
$H_{\rm F}$&      &$\sqrt{(V(\sigma_F)+\Lambda_{\pm})/3}$&   &(\ref{eq:hubblef})\\
$\alpha_F$&      &$FX_\pm/H_{\rm F}$&  &Below (\ref{eq:hubblef})\\
\hline
\end{tabular}
\end{table}

\begin{acknowledgments}
We thank C.~de~Rham, A.~J.~Tolley, R.~Gobbetti,
A.~E.~G\"umr\"uk\c{c}\"uo\u{g}lu, C.~Lin, S.~Mukohyama, K. Sugimura
and T. Tanaka for helpful discussions.
This work was supported in part by the Grant-in-Aid for the Global
COE Program ``The Next Generation of Physics, Spun from Universality
and Emergence'' from the Ministry of Education, Culture, Sports,
Science and Technology (MEXT) of Japan, and by JSPS Grant-in-Aid for
Scientific Research (A) No.~21244033.
RS is supported by a JSPS Grant-in-Aid through the
JSPS postdoctoral fellowship.
We acknowledge the workshop YITP-W-11-26 at Yukawa Institute
for Theoretical Physics, Kyoto University, where
this project was initiated, and we are grateful to the participants of
the workshop YITP-T-12-04 for valuable discussions and comments.
\end{acknowledgments}


\begin{thebibliography}{99}

\bibitem{Fierz:1939}
  M.~Fierz and W.~Pauli,
  Proc.\ Roy.\ Soc.\ Lond.\ A  {\bf 173}, 211-232 (1939).

\bibitem{Boulware:1972}
  D.~G.~Boulware and S.~Deser,
  Phys.\ Rev.\ D  {\bf 6}, 3368-3382 (1972).

\bibitem{Creminelli:2005qk}
  P.~Creminelli, A.~Nicolis, M.~Papucci and E.~Trincherini,
  JHEP {\bf 0509}, 003 (2005).
  [hep-th/0505147].

\bibitem{Rubakov:2008}
  V.~A.~Rubakov and P.~G.~Tinyakov,
  Phys.\ Usp.\ {\bf 51}, 759-792 (2008).
  [arXiv:0802.4379 [hep-th]].

\bibitem{Hinterbichler:2012}
  K.~Hinterbichler,
  Rev.\ Mod.\ Phys.\ {\bf 84}, 671-710 (2012).
  [arXiv:1105.3735 [hep-th]].

\bibitem{Rham:2010}
  C.~de~Rham and G.~Gabadadze,
  Phys.\ Rev.\ D  {\bf 82}, 044020 (2010).
  [arXiv:1007.0443 [hep-th]].


\bibitem{Rham:2011PRL}
  C.~de~Rham, G.~Gabadadze and A.~J.~Tolley,
  Phys.\ Rev.\ Lett.\  {\bf 106}, 231101 (2011).
  [arXiv:1011.1232 [hep-th]].

\bibitem{Hassan:2011vm}
  S.~F.~Hassan and R.~A.~Rosen,
  JHEP {\bf 1107}, 009 (2011).
  [arXiv:1103.6055 [hep-th]].

\bibitem{Hassan:2012}
  S.~F.~Hassan and R.~A.~Rosen,
  Phys.\ Rev.\ Lett.\ {\bf 108}, 041101 (2012).
  [arXiv:1106.3344 [hep-th]].

\bibitem{Hassan:2011tf}
  S.~F.~Hassan, R.~A.~Rosen and A.~Schmidt-May,
  JHEP {\bf 1202}, 026 (2012)
  [arXiv:1109.3230 [hep-th]].

\bibitem{D'Amico:2011jj}
  G.~D'Amico, C.~de Rham, S.~Dubovsky, G.~Gabadadze, D.~Pirtskhalava and A.~J.~Tolley,
  Phys.\ Rev.\ D {\bf 84}, 124046 (2011).
  [arXiv:1108.5231 [hep-th]].

\bibitem{Gumrukcuoglu:2011open}
  A.~E.~G\"umr\"uk\c{c}\"uo\u{g}lu, C.~Lin and S.~Mukohyama,
  JCAP\  {\bf 11}, 030 (2011).
  [arXiv:1109.3845 [hep-th]].

\bibitem{Kobayashi:2012}
  T.~Kobayashi, M.~Siino, M.~Yamaguchi and D.~Yoshida,
  Phys.\ Rev.\ D {\bf 86}, 061505 (2012).
  [arXiv:1205.4938 [hep-th]].

\bibitem{Gratia:2012}
  P.~Gratia, W.~Hu and M.~Wyman,
  Phys.\ Rev.\ D {\bf 86}, 061504(2012).
  [arXiv:1205.4241 [hep-th]].

\bibitem{Koyama:2012prda}
  K.~Koyama, G.~Niz and G.~Tasinato,
   Phys.\ Rev.\ D {\bf 84}, 064033 (2011).
   [arXiv:1104.2143 [hep-th]].

\bibitem{Koyama:2012prl}
  K.~Koyama, G.~Niz and G.~Tasinato,
   Phys.\ Rev.\ Lett.\  {\bf 107}, 131101 (2011).
   [arXiv:1103.4708 [hep-th]].

\bibitem{Langlois:2012hk}
  D.~Langlois and A.~Naruko,
  Class.\ Quant.\ Grav. {\bf 29}, 202001 (2012)
  [arXiv:1206.6810 [hep-th]].


\bibitem{Motohashi:2012jd}
  H.~Motohashi and T.~Suyama,
  Phys.\ Rev.\ D {\bf 86}, 081502 (2012)
  [arXiv:1208.3019 [hep-th]].

\bibitem{Higuchi:1986py}
  A.~Higuchi,
  Nucl.\ Phys.\ B {\bf 282}, 397 (1987).

\bibitem{Grisa:2009yy}
  L.~Grisa and L.~Sorbo,
  Phys.\ Lett.\ B {\bf 686}, 273 (2010).
  [arXiv:0905.3391 [hep-th]].

\bibitem{Fasiello:2012rw}
  M.~Fasiello and A.~J.~Tolley,
  JCAP\  {\bf 1211}, 035 (2012).
  [arXiv:1206.3852 [hep-th]].

\bibitem{Gumrukcuoglu:2011perturb}
  A.~E.~G\"umr\"uk\c{c}\"uo\u{g}lu, C.~Lin and S.~Mukohyama,
  JCAP\  {\bf 1203}, 006 (2012).
  [arXiv:1111.4107 [hep-th]].

\bibitem{Koyama:2011jhep}
 K.~Koyama, G.~Niz and G.~Tasinato,
   JHEP\ {\bf 1112}, 065 (2011).
   [arXiv:1110.2618 [hep-th]].

\bibitem{DeFelice:2012mx}
  A.~De Felice, A.~E.~G\"umr\"uk\c{c}\"uo\u{g}lu and S.~Mukohyama,
  Phys.\ Rev.\ Lett. {\bf 109}, 171101(2012)
  [arXiv:1206.2080 [hep-th]].

\bibitem{Koyama:2012vector}
G.~Tasinato, K.~Koyama and G.~Niz,
   arXiv:1210.3627 [hep-th].

\bibitem{Weinberg:1988cp}
  S.~Weinberg,
  Rev.\ Mod.\ Phys.\  {\bf 61}, 1 (1989).

\bibitem{Nobbenhuis:2004wn}
  S.~Nobbenhuis,
  Found.\ Phys.\  {\bf 36}, 613 (2006).
  [gr-qc/0411093].

\bibitem{Susskind:2003kw}
  L.~Susskind,
  In *Carr, Bernard (ed.): Universe or multiverse?* 247-266
  [hep-th/0302219].

\bibitem{Huang:2012}
  Q.~Huang, Y.~Piao and S.~Zhou,
  Phys.\ Rev.\ D {\bf 86}, 124014 (2012).
  [arXiv:1206.5678 [hep-th]].

\bibitem{Saridakis:2012a}
  E.~N.~Saridakis,
  arXiv:1207.1800 [gr-qc].

\bibitem{Saridakis:2012jcap}
  C.~Cai, C.~Gao and E.~N.~Saridakis,
  JCAP\ {\bf 1210}, 048 (2012)
  [arXiv:1207.3786 [astro-ph]].

\bibitem{Hawking:1981fz}
  S.~W.~Hawking and I.~G.~Moss,
  Phys.\ Lett.\ B {\bf 110}, 35 (1982).

\bibitem{Coleman:1980}
  S.~R.~Coleman and F.~De~Luccia,
  Phys.\ Rev.\ D  {\bf 21}, 3305 (1980).

\bibitem{Tanaka:1992}
  T.~Tanaka and M.~Sasaki,
  Prog.\ Theor.\ Phys.\  {\bf 88}, 503 (1992)

\bibitem{Coleman:1977th}
  S.~R.~Coleman, V.~Glaser and A.~Martin,
  Commun.\ Math.\ Phys.\  {\bf 58}, 211 (1978).

\bibitem{Hassan:2011zd}
  S.~F.~Hassan and R.~A.~Rosen,
  JHEP {\bf 1202}, 126 (2012)
  [arXiv:1109.3515 [hep-th]].

\bibitem{Volkov:2011an}
  M.~S.~Volkov,
  JHEP {\bf 1201}, 035 (2012)
  [arXiv:1110.6153 [hep-th]].

\bibitem{Comelli:2011zm}
  D.~Comelli, M.~Crisostomi, F.~Nesti and L.~Pilo,
  JHEP {\bf 1203}, 067 (2012)
  [Erratum-ibid.\  {\bf 1206}, 020 (2012)]
  [arXiv:1111.1983 [hep-th]].

\bibitem{Hartle:1983}
  J.~B.~Hartle and S.~W.~Hawking,
  Phys.\ Rev.\ D {\bf 28}, 2960 (1983).

\bibitem{Vilenkin:1984}
  A.~Vilenkin,
  Phys.\ Rev.\ D {\bf 30}, 509 (1984).

\bibitem{Hawking:1984}
  S.~W.~Hawking,
  Phys.\ Lett.\ B {\bf 134}, 403 (1984).

\bibitem{Starobinsky:1986fx}
  A.~Starobinsky,
  In *De Vega, H.j. ( Ed.), Sanchez, N. ( Ed.): Field Theory, Quantum Gravity and Strings*, 107-126



\end{thebibliography}
\end{document}